\documentclass[aps,twocolumn,floatfix,superscriptaddress,preprintnumbers,showpacs,showkeys]{revtex4}
\usepackage{amssymb}
\usepackage{amsmath}
\usepackage{amsfonts}
\usepackage{epsfig}
\usepackage{graphicx}
\usepackage{tabularx}
\usepackage{verbatim}
\newcommand{\eq}{\begin{eqnarray}}
\newcommand{\en}{\end{eqnarray}}

\begin{document}

\title{Unified description of BaBar and Belle data on the
bottomonia decays $\Upsilon(mS) \to \Upsilon(ns) \pi^+ \pi^-$}

\author{Yurii S. Surovtsev}
\affiliation{Bogoliubov Laboratory of Theoretical Physics,
Joint Institute for Nuclear Research, 141980 Dubna, Russia}
\author{Petr Byd\v{z}ovsk\'y}
\affiliation{Nuclear Physics Institute of the AS CR, 25068 \v{R}e\v{z},
Czech Republic}
\author{Thomas Gutsche}
\affiliation{Institut f\"ur Theoretische Physik,
Universit\"at T\"ubingen,
Kepler Center for Astro and Particle Physics,
Auf der Morgenstelle 14, D-72076 T\"ubingen, Germany}
\author{Robert~Kami\'nski}
\affiliation{Institute of Nuclear Physics of the PAN, Cracow 31342, Poland}
\author{Valery E. Lyubovitskij}
\affiliation{Institut f\"ur Theoretische Physik,
Universit\"at T\"ubingen,
Kepler Center for Astro and Particle Physics,
Auf der Morgenstelle 14, D-72076 T\"ubingen, Germany}
\affiliation{Department of Physics, Tomsk State University,
634050 Tomsk, Russia}
\affiliation{Mathematical Physics Department,
Tomsk Polytechnic University,
Lenin Avenue 30, 634050 Tomsk, Russia}
\author{Miroslav Nagy}
\affiliation{Institute of Physics, SAS, Bratislava 84511, Slovak Republic}

\date{\today}

\begin{abstract}

We present a unified analysis of the decays of bottomonia
$\Upsilon(mS)\to\Upsilon(nS)\pi\pi$ ($m>n$, $m=2,3,4,5,$ $n=1,2,3$),
charmonia $J/\psi\to\phi(\pi\pi, K\overline{K})$, $\psi(2S)\to J/\psi\pi\pi$, 
and the isoscalar $S$-wave processes $\pi\pi\to\pi\pi,K\overline{K},\eta\eta$.
In this analysis we extend our recent study of low-lying $(m=2,3)$ radial 
excitations of bottomonia to modes involving higher $(m=4,5)$ excited states.
Similarly as for the data on lower radial excitations, we confirm 
that the data for higher 
radially excited states from the {\it BABAR} and Belle collaborations can 
be described under conditions that the final bottomonium is a spectator and 
the multichannel $\pi\pi$
scattering is considered in a model-independent approach based on analyticity,
unitarity and the uniformization procedure.
Indeed we show that the dipion mass distributions in the two-pion transitions
of both charmonia and bottomonia states are explained by a unified mechanism
based on the contribution of the $\pi\pi$ and $K\overline{K}$ coupled channels
including their interference (final-state interactions). 
Therefore, our main result is that the lower and higher radially excited
states of charmonia and bottomonia have no specific features in mutual comparison 
and can be understood in a unified picture, e.g. proposed by our approach.

\end{abstract}



\maketitle

\section{Introduction}

Presently available data on bottomonia decays
$\Upsilon(mS)\to\Upsilon(ns) \pi^+ \pi^-$ ($n=1,2,3$)
extracted by different collaborations (ARGUS~\cite{Argus}, 
CLEO~\cite{CLEO}, CUSB~\cite{CUSB},
Crystal Ball~\cite{Crystal_Ball(85)}, Belle~\cite{Belle}, 
and {\it BABAR}~\cite{BaBar06}) and for both lower and higher radial
excitations of $\Upsilon(mS)$ offer a possibility of their unified theoretical
description. Recently, in Ref.~\cite{SBGKLN-prd15} we focused only on the
decays of lower radial excitation of bottomonia. Restriction to lower radial 
excitation was not specific; therefore we need to extend our analysis 
to higher excited states. 
This paper is devoted to the unified description of {\it BABAR}~\cite{BaBar06}
and Belle~\cite{Belle} data on the decays
$\Upsilon(4S,5S)\to\Upsilon(ns) \pi^+ \pi^-$ ($n=1,2,3$) using the same set of
couplings parametrizing the scattering amplitudes as in our recent 
study~\cite{SBGKLN-prd15},
where we focused on the decays of lower radial excitations of bottomonia.
Now both lower and higher radial excitations of bottomonia are analyzed
in a unified picture using all available data on the two-pion transitions
$\Upsilon(mS)\to\Upsilon(nS)\pi\pi$ ($m>n$, $m=2,3,4,5,$ $n=1,2,3$)
of the $\Upsilon$ mesons~\cite{Argus}-\cite{BaBar06}.
It is important to note  that the analysis of bottomonia
decays has been done together with the isoscalar $S$-wave processes
$\pi\pi\!\to\!\pi\pi,K\overline{K},\eta\eta$ and the charmonium decay 
transitions $J/\psi\to\phi(\pi\pi, K\overline{K})$, 
$\psi(2S)\to J/\psi\pi\pi$ using data
from the Crystal Ball, DM2, Mark~II, Mark~III, 
and BES~II collaborations~\cite{SBLKN-prd14}.
At that, the formalism for the analysis of multichannel $\pi\pi$ scattering 
is based on analyticity, unitarity and the uniformization procedure.

One of the main objectives of our study is to shed some light on the nature of
scalar mesons. The possibility for using the two-pion transitions of heavy 
quarkonia as a laboratory for studying the $f_0$ mesons is related to the 
expectation that the dipion
is produced in a relative $S$ wave whereas the final quarkonium state remains
a spectator~\cite{MP-prd93}.
Many efforts were undertaken to study scalar mesons, mainly by analyzing
multichannel $\pi\pi$ scattering.
The problem of a unique structure interpretation of the scalar mesons
is far away from being solved completely~\cite{PDG-14}.
Previously we analyzed data on the decays of low-lying radial excitations
of bottomonia $\Upsilon(mS)\to\Upsilon(nS)\pi\pi$ ($m>n, m=2,3, n=1,2$), on
multichannel $\pi\pi$ scattering, and on the charmonium decay processes. 
We showed~\cite{SBGKLN-prd15} that the considered bottomonia decay data
do not really offer new insights into the nature of the scalar mesons 
which were not already deduced in previous analyses of 
pseudoscalar-meson-scattering processes.
The results of the analysis have confirmed all our earlier conclusions
on the scalar mesons~\cite{SBLKN-prd14}. However, the problem must be 
considered further by allowing for an extended analysis including  available 
data on the $\Upsilon(4S,5S)$ decays.

Note that the previous analysis of the process 
$\Upsilon(3S)\to\Upsilon(1S)\pi\pi$
has already given us an opportunity to obtain interesting conclusions on
the mechanism of this decay~\cite{SBGKLN-prd15}, which is able to explain
the enigmatic two-humped shape of the dipion mass distribution.
This distribution might be the result of the destructive interference of the
relevant contributions to the decay $\Upsilon(3S)\to\Upsilon(1S)\pi\pi$.
However, in this scenario the phase space cuts off possible contributions,
which might interfere destructively with the $\pi\pi$-scattering contribution
giving the specific shape of the dipion spectrum. In a number of works
(see, e.g., Ref.~\cite{SimVes} and the references therein, and our discussion
in Ref.~\cite{SBGKLN-prd15}) various (sometimes rather doubtful) assumptions 
were made to obtain the needed result. We have explained this effect on the 
basis of our previous conclusions without any additional assumptions.
In Refs.~\cite{SBLKN-prd14,SBLKN-jpgnpp14,SBKLN-PRD12} 
we have shown the following:
if a wide resonance cannot decay into a channel which opens above its mass,
and if the resonance is strongly coupled to this channel [e.g. $f_0(500)$ and
the $K\overline{K}$ channel], then this resonance should be treated as
a multichannel state. The closed channel should be included while taking
into account the Riemann-surface sheets related to the threshold branch point
of this channel and performing the combined analysis of the coupled channels.

In the present extension we include the $\Upsilon(4S)$ and $\Upsilon(5S)$
which are distinguished from the lower $\Upsilon$ states by the fact that
their masses are above the $B\overline{B}$ thresholds.
These higher states predominantly decay into pairs of the $B$-meson family
because these modes are not suppressed by the OZI rule: the $\Upsilon(4S)$
decays into $B\overline{B}$ pairs form more than $96$\% of the total width; 
for the $\Upsilon(5S)$ these decay modes make up about $90$\%. 
Therefore, there naturally appears a desire to use this fact in explaining 
the characteristic shape of the dipion mass distribution in the decays 
$\Upsilon(mS)\to\Upsilon(nS)\pi\pi$ ($m>n$, $m=2,3,4,5,$ $n=1,2,3$). 
E.g., in Ref.~\cite{Chen_Liu_Li-epjc11}, one supposed that a pion pair is formed 
in the $\Upsilon(4S)$ decay both as a direct production and as the sequential process
$\Upsilon(4S)\to B\overline{B}\to\Upsilon(nS)+f_0\to\Upsilon(nS)+\pi^+\pi^-$ ($n=1,2$). 
Though an allowance for contributions of these two mechanisms with a relative phase 
reproduces satisfactorily the data on decays $\Upsilon(4S)\to\Upsilon(2S,1S)\pi^+\pi^-$, 
it seems that the former assumption is not reasonable because the pions interact strongly.

In contrast to the very big contributions to the total widths of the $\Upsilon(4S,5S)$ 
from decays into pairs of the $B$-meson family, the processes of interest are strongly 
reduced decay modes: the decays $\Upsilon(4S)\to\Upsilon(1S)\pi\pi$ and 
$\Upsilon(4S)\to\Upsilon(2S)\pi\pi$ form  about $(8.1\pm 0.6)\times 10^{-5}$\% and 
$(8.6\pm1.3)\times 10^{-5}$\% of the total width, 
and $\Upsilon(5S)\to\Upsilon(1S,2S,3S)\pi\pi$, $(5\div8)\times 10^{-3}$\%~\cite{PDG-14}. 
The total widths of $\Upsilon(5S)$ and $\Upsilon(4S)$ are 110 and 20.5~MeV, 
respectively, and the one of the $\Upsilon(3S)$ (on which we already have clarified 
the mechanism of the two-pion transitions \cite{SBGKLN-prd15}) is 20.32~keV. 
The partial decay widths of $\Upsilon(5S)\to\Upsilon(1S,2S,3S)\pi\pi$ are almost 
of the same order as the ones of the decays $\Upsilon(3S)\to\Upsilon(1S,2S)\pi\pi$. 
The decay widths of $\Upsilon(4S)\to\Upsilon(1S,2S)\pi\pi$ are even smaller than 
the latter ones by about 2 orders of magnitude.

The above comparison of decay widths implies that in the two-pion transitions
of $\Upsilon(4S)$ and $\Upsilon(5S)$ the basic mechanism, which explains the
dipion mass distributions, cannot be related to the $B\bar B$ transition
dynamics.
We shall show that the two-pion transitions both of bottomonia and
charmonia are explained by a unified mechanism. It is based on our previous
conclusions on the wide resonances \cite{SBLKN-prd14,SBLKN-jpgnpp14,SBKLN-PRD12}
and is related to the interference of the contributions of multichannel $\pi\pi$
scattering in the final-state interaction.

We also work out the role of the individual $f_0$ resonances in contributing to
the dipion mass distributions in the decays
$\Upsilon(4S,5S)\to\Upsilon(nS) \pi^+ \pi^-$ ($n=1,2,3$).
For this purpose, in the Appendix, we summarize and discuss some formulas and
results from our previous paper~\cite{SBLKN-prd14}.

\section{Multichannel $\pi\pi$ scattering in two-pion transitions of bottomonia}

When carrying out our analysis, data for the processes
$\pi\pi\to\pi\pi,K\overline{K},\eta\eta$ are taken from many sources
(see the corresponding references in~\cite{SBLKN-prd14}).
Formalism for analyzing the multichannel $\pi\pi$ scattering
is presented briefly in the Appendix.
The combined analysis including decay data on $J/\psi\to\phi(\pi\pi,K\overline{K})$
and $\psi(2S)\to J/\psi\pi\pi$ from the Mark~III, DM2, BES~II, Mark~II, and Crystal Ball
collaborations (see corresponding references also in~\cite{SBLKN-prd14}) was found
to be important for getting unique solutions to the $f_0$-meson parameters:
first we solved the ambiguity in the parameters of the $f_0(500)$~\cite{SBL-prd12}
in favor of the wider state; second, the parameters of the other $f_0$ mesons had
small corrections~\cite{SBLKN-prd14}. A further addition of decay data on
$\Upsilon(mS)\to\Upsilon(nS)\pi\pi$ ($m>n, m=2,3, n=1,2$) from ARGUS, CLEO, CUSB, and
the Crystal Ball collaborations in a combined analysis did not add any new constraints
on the $f_0$ mesons, thus confirming the previous conclusions about these states. 
However, the analysis resulted in an interesting explanation of the enigmatic two-humped
shape of the dipion spectrum in the decay $\Upsilon(3S)\to\Upsilon(1S)\pi\pi$:
this shape is proved to be stipulated by a destructive interference of the $\pi\pi$
and $K\overline{K}$ coupled-channel contributions to the final state of
this decay~\cite{SBGKLN-prd15}.

In the present manuscript we further extend the analysis of the two-pion transitions
of radially excited $\Upsilon$ mesons to higher states --- $\Upsilon(4S)$ and $\Upsilon(5S)$.
The used formalism for calculating the dimeson mass distributions in the $\Upsilon(mS)$
decays is analogous to the one proposed in Ref.~\cite{MP-prd93} for the decays
$J/\psi\to\phi(\pi\pi, K\overline{K})$ and $V^{\prime}\to V\pi\pi$ ($V=\psi,\Upsilon$).
I.e., it was assumed that the pion pairs in the final state have zero isospin and spin. 
Only these pairs of pions undergo final-state interactions whereas the final $\Upsilon(nS)$ 
meson ($n<m$) acts as a spectator. 
This decay model is justified by the consideration of quark diagrams for the processes of 
interest and by allowance for the fact that we deal with the two-pion transitions of the 
radially excited states to lower ones of the same family; therefore, it is reasonable to 
expect that the dipion is produced in a relative $S$ wave and the final bottomonium state 
remains the spectator.
The amplitudes for the decays $\Upsilon(mS)\to\Upsilon(nS)\pi\pi$ 
($m>n$, $m=2,3,4,5,$ $n=1,2,3$) include the scattering
amplitudes $T_{ij}$ $(i,j=1-\pi\pi,2-K\overline{K})$ as follows: 
\begin{equation}
F_{mn}(s) = (\rho_{mn}^0+\rho_{mn}^1\,s)\,T_{11}
+ (\omega_{mn}^0+\omega_{mn}^1\,s)\,T_{21},
\end{equation}
where indices $m$ and $n$ correspond to $\Upsilon(mS)$ and $\Upsilon(nS)$, respectively.
The free parameters $\rho_{mn}^0$, $\rho_{mn}^1$, $\omega_{mn}^0$, and $\omega_{mn}^1$
depend on the couplings of the $\Upsilon(mS)$ to the channels $\pi\pi$ and $K\overline{K}$.
The model-independent amplitudes $T_{ij}$ are expressed through the $S$-matrix elements
shown in the Appendix 
\eq
S_{ij}=\delta_{ij}+2i\sqrt{\rho_1\rho_2}\, T_{ij}\,, 
\en
where $\rho_i=\sqrt{1-s_i/s}$ and $s_i$ is the reaction threshold.
The expressions for the dipion mass distributions in the decay
$\Upsilon(mS)\to\Upsilon(nS)\pi\pi$ are
\eq
N|F|^{2}\sqrt{(s-s_1) \lambda(m_{\Upsilon(mS)}^2,s,m_{\Upsilon(nS)}^2)}\,,
\en
where $\lambda(x,y,z)=x^2+y^2+z^2-2xy-2yz-2xz$ is the K\"allen function.
The normalization constants $N$ are  determined by a fit to the specific
experiment and collected in Table~\ref{tab:Normlazations}.
Parameters of the coupling functions of the decay particles
$\Upsilon(mS)~(m=2,...,5)$ to channel~$i$ 
obtained in the analysis are shown in Tables~\ref{tab:constants_rho}
and~\ref{tab:constants_omega}.
A satisfactory combined description of all considered processes is obtained
with a total $\chi^2/\mbox{ndf}=824.236/(714-91)\approx1.32$.
The $\chi^2/\mbox{ndp}$ (ndp is number of data points) 
estimates for the processes 
$\pi\pi\to\pi\pi,K\overline{K},\eta\eta$, and specific decay modes are
collected in Table~\ref{tab:chi2}.

\begin{center}
\begin{table}
\caption{Normalization constants $N$. \label{tab:Normlazations}}
\vspace*{0.1cm}
\def\arraystretch{1.5}
\begin{tabular}{ccl}
\hline\hline
Process & $ \ \ \ N \ \ \ $ & Collaboration \\
$\Upsilon(2S) \to \Upsilon(1S)\pi^+\pi^-$ & 4.3439 & ARGUS~\cite{Argus} \\
\hline
$\Upsilon(2S) \to \Upsilon(1S)\pi^+\pi^-$ & 2.1776 & CLEO~\cite{CLEO} \\
\hline
$\Upsilon(2S) \to \Upsilon(1S)\pi^+\pi^-$ & 1.2011 & CUSB~\cite{CUSB} \\
\hline
$\Upsilon(2S) \to \Upsilon(1S)\pi^0\pi^0$ & 0.0788 & Crystal Ball~\cite{Crystal_Ball(85)} \\
\hline
$\Upsilon(3S) \to \Upsilon(1S)\pi^+\pi^-$ & 0.5096 & CLEO~\cite{CLEO07} \\
\hline
$\Upsilon(3S) \to \Upsilon(1S)\pi^0\pi^0$ & 0.2235 & CLEO~\cite{CLEO07} \\
\hline
$\Upsilon(3S) \to \Upsilon(2S)\pi^+\pi^-$ & 7.7397 & CLEO~\cite{CLEO(94)} \\
\hline
$\Upsilon(3S) \to \Upsilon(2S)\pi^0\pi^0$ & 3.8587 & CLEO~\cite{CLEO(94)} \\
\hline
$\Upsilon(4S) \to \Upsilon(1S)\pi^+\pi^-$ & 7.1476 & {\it BABAR}~\cite{BaBar06} \\
\hline
$\Upsilon(4S) \to \Upsilon(1S)\pi^+\pi^-$ & 0.5553 & Belle~\cite{Belle} \\
\hline
$\Upsilon(4S) \to \Upsilon(2S)\pi^+\pi^-$ & 58.143 & {\it BABAR}~\cite{BaBar06} \\
\hline
$\Upsilon(5S) \to \Upsilon(1S)\pi^+\pi^-$ & 0.1626 & Belle~\cite{Belle} \\
\hline
$\Upsilon(5S) \to \Upsilon(2S)\pi^+\pi^-$ & 4.8355 & Belle~\cite{Belle} \\
\hline
$\Upsilon(5S) \to \Upsilon(3S)\pi^+\pi^-$ & 10.858 & Belle~\cite{Belle} \\
\hline\hline
\end{tabular}

\caption{Parameters of the coupling functions $\rho_{ij}^k$. \label{tab:constants_rho}}
\vspace*{0.1cm}
\def\arraystretch{1.5}
\begin{tabular}{cc|cc}
\hline\hline
Parameter & Numerical value & Parameter & Numerical value \\
\hline
$\rho^{0}_{21}$ & 0.4050 & $\rho^{1}_{21}$ & 47.0963 \\
\hline
$\rho^{0}_{31}$ & 1.0827 & $\rho^{1}_{31}$ & $-$2.7546 \\
\hline
$\rho^{0}_{32}$ & 7.3875 & $\rho^{1}_{32}$ & $-$2.5598 \\
\hline
$\rho^{0}_{41}$ & 0.6162 & $\rho^{1}_{41}$ & $-$2.5715 \\
\hline
$\rho^{0}_{42}$ & 2.3290 & $\rho^{1}_{42}$ & $-$7.3511 \\
\hline
$\rho^{0}_{51}$ & 0.7078 & $\rho^{1}_{51}$ & 4.0132  \\
\hline
$\rho^{0}_{52}$ & 0.8133 & $\rho^{1}_{52}$ & 2.2061  \\
\hline
$\rho^{0}_{53}$ & 0.8946 & $\rho^{1}_{53}$ & 2.5380  \\
\hline\hline
\end{tabular}

\caption{Parameters of the coupling functions
$\omega_{ij}^k$. \label{tab:constants_omega}}
\vspace*{0.1cm}
\def\arraystretch{1.5}
\begin{tabular}{cc|cc}
\hline\hline
Parameter & Numerical value & Parameter & Numerical value \\
\hline
$\omega_{21}^0$ & 1.3352 & $\omega_{21}^1$ & $-$21.4343 \\
\hline
$\omega_{31}^0$ & 0.8615 & $\omega_{31}^1$ & 0.6600   \\
\hline
$\omega_{32}^0$ & 0.0 & $\omega_{31}^1$ & 0.0   \\
\hline
$\omega_{41}^0$ & -0.8467 & $\omega_{41}^1$ & 0.2128  \\
\hline
$\omega_{42}^0$ & 1.8096  & $\omega_{42}^1$ & $-$10.1477  \\
\hline
$\omega_{51}^0$ & 4.8380  & $\omega_{51}^1$ & $-$3.9091  \\
\hline
$\omega_{52}^0$ & -0.7973 & $\omega_{52}^1$ & 0.3247  \\
\hline
$\omega_{53}^0$ & 0.6270  & $\omega_{51}^1$ & $-$0.0483  \\
\hline\hline
\end{tabular}
\end{table}
\end{center}

\begin{center}
\begin{table}
\caption{$\chi^2/\mbox{ndp}$ estimates for specific decay modes.
\label{tab:chi2}}
\vspace*{0.1cm}
\def\arraystretch{1.5}
\begin{tabular}{lc}
\hline\hline
Process & $\chi^2/\mbox{ndp}$ \\
\hline
$\pi\pi$ scattering & 0.90 \\
\hline
$\pi\pi\to K\overline{K}$ & 1.16 \\
\hline
$\pi\pi\to\eta\eta$ & 0.87 \\
\hline
$J/\psi\to\phi(\pi^+\pi^-, K^+K^-)$ & 1.36 \\
\hline
$\psi(2S)\to J/\psi(\pi^+\pi^-,\pi^0\pi^0)$ & 2.43 \\
\hline
$\Upsilon(2S)\to\Upsilon(1S)(\pi^+\pi^-,\pi^0\pi^0)$ & 1.01 \\
\hline
$\Upsilon(3S)\to\Upsilon(1S)(\pi^+\pi^-,\pi^0\pi^0)$ & 0.67 \\
\hline
$\Upsilon(3S)\to\Upsilon(2S)(\pi^+\pi^-,\pi^0\pi^0)$ & 0.61 \\
\hline
$\Upsilon(4S)\to\Upsilon(1S)(\pi^+\pi^-)$ & 0.27 \\
\hline
$\Upsilon(4S)\to\Upsilon(2S)(\pi^+\pi^-)$ & 0.27 \\
\hline
$\Upsilon(5S)\to\Upsilon(1S)(\pi^+\pi^-)$ & 1.80 \\
\hline
$\Upsilon(5S)\to\Upsilon(2S)(\pi^+\pi^-)$ & 1.08 \\
\hline
$\Upsilon(5S)\to\Upsilon(3S)(\pi^+\pi^-)$ & 0.81 \\
\hline\hline
\end{tabular}

\caption{Background parameters for the minimal
set of scalar mesons $f_0(500)$, $f_0(980)$
and $f_0'(1500)$. \label{tab:back1}}
\vspace*{0.1cm}
\def\arraystretch{1.5}
\begin{tabular}{ccc}
\hline\hline
$a_{11}$ & $a_{1\sigma}$ & $a_{1v}$ \\
0.0        & 0.0321        & 0.0        \\
\hline
$b_{11}$ & $b_{1\sigma}$ & $b_{1v}$ \\
$-$0.0051  & 0.0             & 0.04     \\
\hline
$a_{21}$ & $a_{2\sigma}$ & $a_{2v}$ \\
$-$1.6425  & $-$0.3907       & $-$7.274   \\
\hline
$b_{21}$ & $b_{2\sigma}$ & $b_{2v}$ \\
0.1189   & 0.2741        & 5.823    \\
\hline
$b_{31}$ & $b_{3\sigma}$ & $b_{3v}$ \\
0.7711   & 0.505         & 0.0   \\
\hline\hline
\end{tabular}

\caption{Background parameters for the set
of scalar mesons when the $f_0(500)$ is 
switched off. \label{tab:back2}}
\vspace*{0.1cm}
\def\arraystretch{1.5}
\begin{tabular}{ccc}
\hline\hline
$a_{11}$ & $a_{1\sigma}$ & $a_{1v}$ \\
0.3513   & $-$0.2055       & 0.207    \\
\hline
$b_{11}$ & $b_{1\sigma}$ & $b_{1v}$ \\
$-$0.0077  & 0.0             & 0.0378   \\
\hline
$a_{21}$ & $a_{2\sigma}$ & $a_{2v}$ \\
$-$1.8597  & 0.1688       & $-$7.519    \\
\hline
$b_{21}$ & $b_{2\sigma}$ & $b_{2v}$ \\
0.161    & 0.0             & 6.94     \\
\hline
$b_{31}$ & $b_{3\sigma}$ & $b_{3v}$ \\
0.7758   & 0.4985        & 0.0   \\
\hline\hline
\end{tabular}
\end{table}
\end{center}

In Figs.~1 and 2 we show the fits (solid lines) to the experimental data
of the {\it BABAR}~\cite{BaBar06} and Belle~\cite{Belle} collaborations 
on the bottomonia
decays --- $\Upsilon(4S,5S)\to\Upsilon(nS) \pi^+ \pi^-$ ($n=1,2,3$) ---
in the combined analysis with the lower bottomonia decays ---
$\Upsilon(mS)\to\Upsilon(nS)\pi\pi$ ($m>n, m=2,3, n=1,2$) ---
with the processes $\pi\pi\to\pi\pi,K\overline{K},\eta\eta$,n
and the charmonia decays ---
$J/\psi\to\phi(\pi\pi, K\overline{K})$, $\psi(2S)\to J/\psi\pi\pi$.
The curves demonstrate an interesting behavior ---
a bell-shaped form in the near-$\pi\pi$-threshold region [especially
for the $\Upsilon(4S)\to\Upsilon(2S) \pi^+ \pi^-$],
smooth dips near a dipion mass of 0.6~GeV in $\Upsilon(4S,5S)\to\Upsilon(1S) \pi^+ \pi^-$
and of about 0.44~GeV in $\Upsilon(4S)\to\Upsilon(2S) \pi^+ \pi^-$, and
sharp dips of about 1~GeV in the $\Upsilon(4S,5S)\to\Upsilon(1S) \pi^+ \pi^-$ transition.
This shape of the dipion mass distribution is obviously explained by
the interference between the $\pi\pi$-scattering and $K\overline{K}\to\pi\pi$
contributions to the final states of these decays --- by the constructive interference
in the near-$\pi\pi$-threshold region and by a destructive one in the dip regions.
Whereas the data on $\Upsilon(5S)\to\Upsilon(1S) \pi^+ \pi^-$ confirm
the sharp dips near 1~GeV, the scarce data on $\Upsilon(4S)\to\Upsilon(1S) \pi^+ \pi^-$
do not allow for such a unique conclusion yet.
We further investigated the role of the individual $f_0$ resonances
in contributing to the shape of the dipion mass distributions in the decays
$\Upsilon(4S,5S)\to\Upsilon(nS) \pi^+ \pi^-$ ($n=1,2,3$).
In this case we switched off only those resonances [$f_0(500)$, $f_0(1370)$,
$f_0(1500)$ and $f_0(1710)$], removal of which can be somehow compensated by
correcting the background (maybe, with elements of the pseudobackground) to have
the more-or-less acceptable description of the multichannel $\pi\pi$ scattering.

First, when leaving out before-mentioned resonances, a minimal set of the $f_0$
mesons consisting of the $f_0(500)$, $f_0(980)$, and $f_0^\prime(1500)$ is
sufficient to achieve a description of the processes
$\pi\pi\!\to\!\pi\pi,K\overline{K},\eta\eta$ with a total $\chi^2/\mbox{ndf}\approx1.20$.
The obtained, adjusted background parameters are shown in Table~\ref{tab:back1}.

Second, from these three mesons only the $f_0(500)$ can be switched off while still
obtaining a reasonable description of multichannel $\pi\pi$ scattering (though with
an appearance of the pseudobackground) with a total $\chi^2/\mbox{ndf}\approx1.43$
and with the corrected background parameters, which are shown in Table~\ref{tab:back2}.

\begin{figure}[!htb]
\begin{center}
\includegraphics[width=0.48\textwidth]{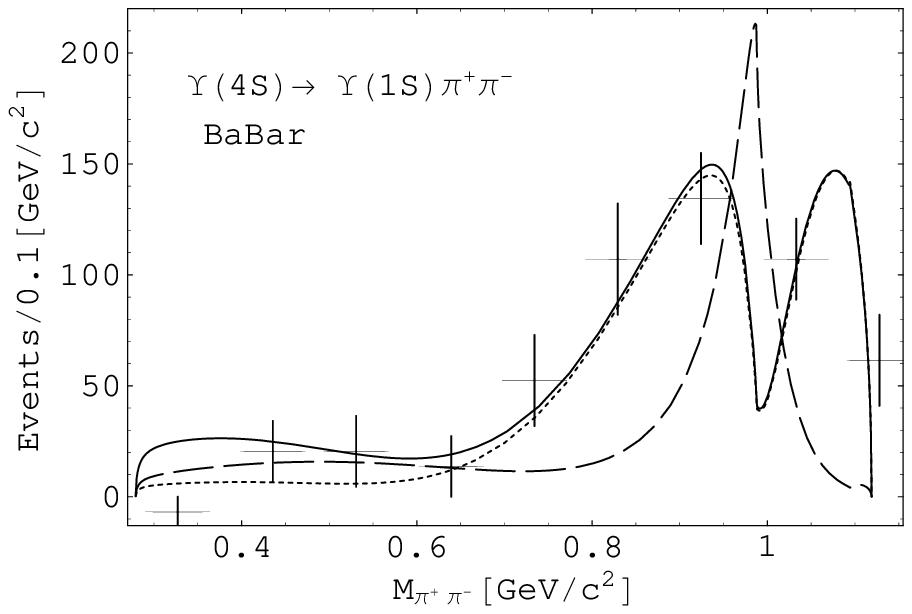}\\
\includegraphics[width=0.48\textwidth]{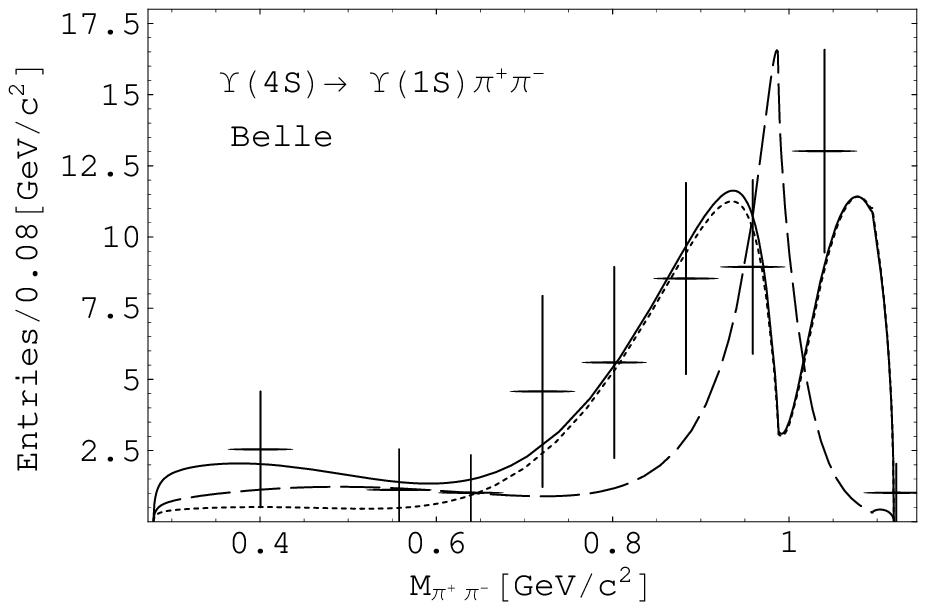}\\
\includegraphics[width=0.48\textwidth]{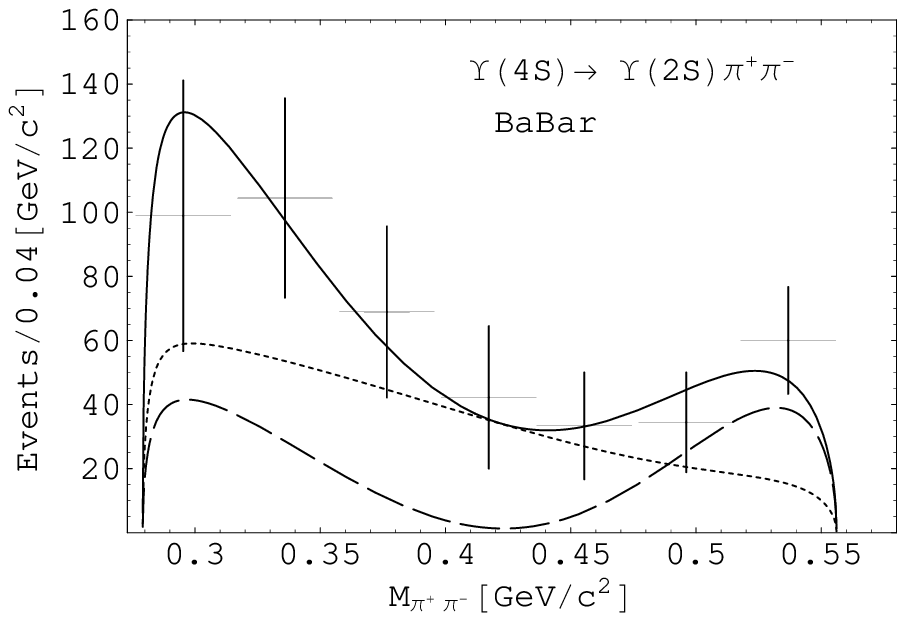}
\vspace*{-0.17cm}\caption{
The decays $\Upsilon(4S)\to\Upsilon(1S)\pi\pi$ and $\Upsilon(4S)\to\Upsilon(2S)\pi\pi$.
The solid lines correspond to the contribution of all relevant resonances;
the dotted, of the $f_0(500)$, $f_0(980)$, and $f_0^\prime(1500)$;
the dashed, of the $f_0(980)$ and $f_0^\prime(1500)$.}
\end{center}
\end{figure}
\begin{figure}[!htb]
\begin{center}
\includegraphics[width=0.48\textwidth]{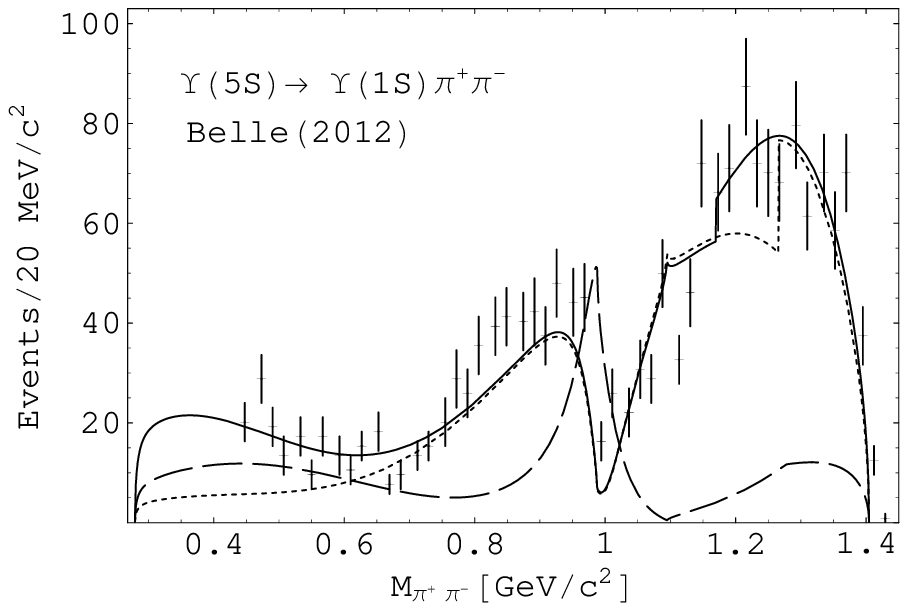}
\includegraphics[width=0.48\textwidth]{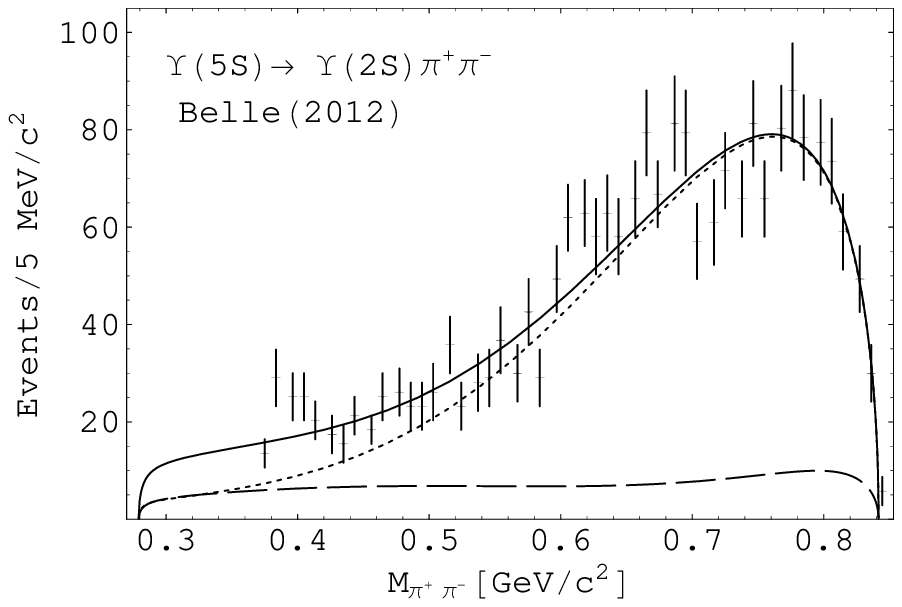}\\
\includegraphics[width=0.48\textwidth]{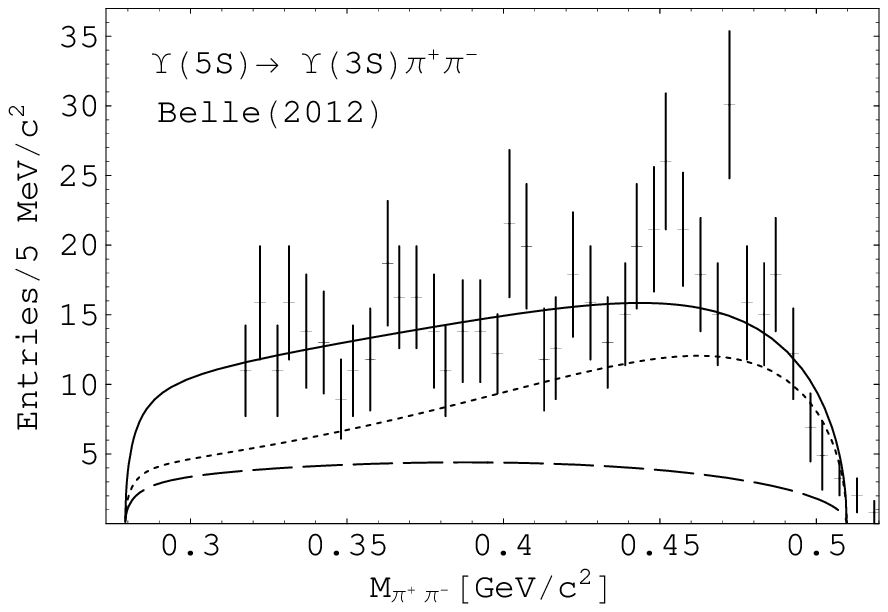}
\vspace*{-0.17cm}\caption{
The decays $\Upsilon(5S)\to\Upsilon(nS)\pi^+ \pi^-$ ($n=1,2,3$).
The solid lines correspond to the contribution of all relevant resonances;
the dotted, of the $f_0(500)$, $f_0(980)$, and $f_0^\prime(1500)$;
the dashed, of the $f_0(980)$ and $f_0^\prime(1500)$.}
\end{center}
\end{figure}

In Figs.~1 and 2 variants of the calculations of the dipion mass distributions
with contributions from the $f_0(500)$, $f_0(980)$, and $f_0^\prime(1500)$ and
from the $f_0(980)$, and $f_0^\prime(1500)$ are shown by the dotted and dashed lines,
respectively. It is seen that the sharp dips near 1~GeV in the $\Upsilon(4S,5S)$
decays are related to the $f_0(500)$ contribution in the interfering amplitudes
of $\pi\pi$ scattering and the $K\overline{K}\to\pi\pi$ process.

One should also note the unexpected result --- a considerable contribution of
the $f_0(1370)$ to the bell-shaped form in the near-$\pi\pi$-threshold region,
especially in the decay $\Upsilon(4S)\to\Upsilon(2S)\pi\pi$. This is interesting
because the $f_0(1370)$ is predominantly the $s{\bar s}$ state according to the
earlier analysis~\cite{SBLKN-prd14} and practically does not contribute to the
$\pi\pi$-scattering amplitude. However, this state influences noticeably the
$K\overline{K}$ scattering; e.g., it was shown that the $K\overline{K}$-scattering
length is very sensitive to whether this state exists or not~\cite{SKN-epja02}.

\section{Summary}

We performed a combined analysis of data on isoscalar $S$-wave processes
$\pi\pi\to\pi\pi,K\overline{K},\eta\eta$, on the decays of the charmonia ---
$J/\psi\to\phi(\pi\pi, K\overline{K})$, $\psi(2S)\to J/\psi\,\pi\pi$ ---
and of the bottomonia ---
$\Upsilon(mS)\to\Upsilon(nS)\pi\pi$ ($m>n$, $m=2,3,4,5,$ $n=1,2,3$)
from the ARGUS, Crystal Ball, CLEO, CUSB, DM2, Mark~II, Mark~III, BES~II, {\it BABAR},
and Belle collaborations.
It is interesting that the expansion of the
analyzed data by adding the ones on the above bottomonia decays does not change
practically the values of the fitted resonance and background parameters in comparison
with the combined analysis only of the above multichannel $\pi\pi$ scattering and
charmonia decays. Therefore, it is possible and reasonable to consider that these
parameters are fixed by the latter.

Here we specifically focused on the unified description of {\it BABAR}~\cite{BaBar06} and
Belle~\cite{Belle} data on the decays $\Upsilon(4S,5S)\to\Upsilon(nS) \pi^+ \pi^-$
($n=1,2,3$).
It was shown that the dipion mass distributions in the two-pion transitions both
of charmonia and bottomonia are explained by a unified mechanism related to
contributions of the $\pi\pi$ and $K\overline{K}$ coupled channels and their
interference. The role of the individual $f_0$ resonances in making up the shape
of the dipion mass distributions in these decays was considered.

When describing the bottomonia decays, we did not change the resonance parameters in
comparison with the ones obtained in the combined analysis of the processes
$\pi\pi\to\pi\pi,K\overline{K},\eta\eta$, and
charmonia decays~\cite{SBLKN-prd14,SBGLKN-npbps13}.
Thus, the results of the analysis confirmed all of our earlier conclusions
on the scalar mesons~\cite{SBLKN-prd14}.

\begin{acknowledgments}

This work was supported in part by the Heisenberg-Landau Program,
the Votruba-Blokhintsev Program for Cooperation of Czech Republic
with JINR, the Grant Agency of the Czech Republic (Grant No. P203/15/04301),
the Grant Program of Plenipotentiary of Slovak Republic at JINR,
the Bogoliubov-Infeld Program for Cooperation of Poland with JINR,
the Tomsk State University Competitiveness Improvement Program, 
the Russian Federation program ``Nauka'' (Contract No. 0.1526.2015, 3854), 
Slovak Grant Agency VEGA under Contract No. 2/0197/14,
and by the Polish National Science Center (NCN) Grant No. DEC-2013/09/B/ST2/04382.

\end{acknowledgments}

\appendix\section{The model-independent amplitudes for multi-channel $\pi\pi$ scattering}

Considering multichannel $\pi\pi$ scattering, we shall deal with the three-channel
case (namely with $\pi\pi\!\to\!\pi\pi,K\overline{K},\eta\eta$) because it was
shown~\cite{SBLKN-jpgnpp14,SBKLN-PRD12} that this is a minimal number of coupled
channels needed for obtaining reasonable and correct values of the
scalar-isoscalar resonance parameters.

The three-channel $S$ matrix is determined on the eight-sheeted Riemann surface.
The matrix elements $S_{ij}$, where $i,j=1,2,3$ denote the channels, have
right-hand cuts along the real axis of the complex $s$ plane ($s$ is the
invariant total energy squared); starting with the channel thresholds
$s_i$ ($i=1,2,3$), the left-hand cuts are related to the crossed channels.
The Riemann-surface sheets are numbered according to the signs of the analytic
continuations of the quantities $\sqrt{s-s_i}$ as follows:
$\mbox{signs}\Big(\mbox{Im}\sqrt{s-s_1},~~\mbox{Im}\sqrt{s-s_2},~~{\mbox{Im}}
\sqrt{s-s_3}\Big)~=~+++,\\-++,~--+,~+-+,~+--,~---,~-+-,~++-$
correspond to sheets I, II, $\cdots$, VIII, respectively.

The Riemann-surface structure can be represented by taking the
following uniformizing variable \cite{SBL-prd12} where we have neglected the
$\pi\pi$-threshold branch point and included the $K\overline{K}$- and
$\eta\eta$-threshold branch points and the left-hand branch point at $s=0$ related
to the crossed channels with
\eq \label{w}
w=\frac{\sqrt{(s-s_2)s_3} + \sqrt{(s-s_3)s_2}}{\sqrt{s(s_3-s_2)}}\,, 
\en
where $s_2=4m_K^2$ and $s_3=4m_\eta^2$.
Resonance representations on the Riemann surface are obtained using formulas
from~\cite{SBL-prd12}. Analytic continuations of the $S$-matrix elements
to all sheets are expressed in terms of those on the physical (I) sheet that have
only the resonance zeros (beyond the real axis), at least around the physical
region. Then multichannel resonances are classified. For analytic continuations
the resonance poles on sheets II, IV, and VIII, which
are not shifted due to the coupling of channels, correspond to zeros on the physical
sheet in $S_{11}$, $S_{22}$ and $S_{33}$, respectively. They are at the same
points on the energy plane as the resonance poles (for more details see
Ref.~\cite{SBL-prd12}). It is convenient to classify multichannel resonances
according to resonance zeros on sheet I. In the three-channel case there are
{\it seven types} of resonances corresponding to seven possible situations when
there are resonance zeros on sheet I only in $S_{11}$ -- ({\bf a}), ~~$S_{22}$
-- ({\bf b}), ~~$S_{33}$ -- ({\bf c}), ~~$S_{11}$ and $S_{22}$ -- ({\bf d}), 
~~$S_{22}$ and $S_{33}$ -- ({\bf e}), ~~$S_{11}$ and $S_{33}$ -- ({\bf f}),
~~$S_{11}$, $S_{22}$, and $S_{33}$ -- ({\bf g}). The resonance of every type is
represented by a pair of complex-conjugate {\it clusters} (of poles and
zeros on the Riemann surface).

The $S$-matrix elements $S_{ij}$ are parametrized using the Le Couteur--Newton
relations~\cite{LeCou}. They express the $S$-matrix elements of all coupled processes
in terms of the Jost determinant $d(\sqrt{s-s_1},\cdots,\sqrt{s-s_n})$ which
is a real analytic function with the only square-root branch points at
$\sqrt{s-s_i}=0$.
On the $w$ plane, the Le Couteur--Newton relations are~\cite{SBL-prd12}
\begin{eqnarray}
\label{w:LeCouteur-Newton}
&&S_{11}\!=\!\frac{d^* (-w^*)}{d(w)},~
S_{22}\!=\!\frac{d(-w^{-1})}{d(w)},~ S_{33}\!=\!\frac{d(w^{-1})}{d(w)},\\
&&S_{11}S_{22}-S_{12}^2\!=\!\frac{d^*({w^*}^{-1})}{d(w)},~
S_{11}S_{33}-S_{13}^2\!=\!\frac{d^* (-{w^*}^{-1})}{d(w)}\nonumber
\end{eqnarray}
where now $d(w)$ is free from any branch points.
The $S$-matrix elements are taken as the products $S=S_B S_{res}$;
the main ({\it model-independent}) contribution of resonances given by the
pole clusters is included in the resonance part $S_{res}$; possible remaining small
({\it model-dependent}) contributions of resonances and the influence of channels
which are not taken explicitly into account in the uniformizing variable are included
in the background part $S_B$. The $d_{res}(w)$ function for the resonance part,
which now is free from any branch points, is taken as
\begin{equation}\label{dw_res}
d_{res}(w)=w^{-\frac{M}{2}}\prod_{r=1}^{M}(w+w_{r}^*)\,, 
\end{equation}
where $M$ is the number of resonance zeros. For the background part we have
\begin{equation} \label{bg}
d_B=\mbox{exp}[-i\sum_{n=1}^{3}\frac{\sqrt{s-s_n}}{2m_n}(\alpha_n+i\beta_n)]
\end{equation}
with
\eq
&&\alpha_n=a_{n1}+a_{n\sigma}\frac{s-s_\sigma}{s_\sigma}\theta(s-s_\sigma)+
a_{nv}\frac{s-s_v}{s_v}\theta(s-s_v),\nonumber\\
&&\beta_n=b_{n1}+b_{n\sigma}\frac{s-s_\sigma}{s_\sigma}\theta(s-s_\sigma)+
b_{nv}\frac{s-s_v}{s_v}\theta(s-s_v)\nonumber
\en
where $s_\sigma$ is the $\sigma\sigma$ threshold, $s_v$ the combined threshold
of the $\eta\eta^{\prime},~\rho\rho,~\omega\omega$ channels, which were obtained
in the analysis.
The resonance zeros $w_{r}$ and the background parameters were fixed by fitting
to the data on $\pi\pi\to\pi\pi,K\overline{K},\eta\eta$, and
the charmonium decay processes --- $J/\psi\to\phi(\pi\pi, K\overline{K})$,
$\psi(2S)\to J/\psi\pi\pi$~\cite{SBLKN-prd14}.

The preferred scenario found is when the $f_0(500)$ is described by
the cluster of type ({\bf a}), the $f_0(1370)$, $f_0(1500)$, and $f_0(1710)$ with
type ({\bf c}), and $f_0^\prime(1500)$ by type ({\bf g}); the $f_0(980)$ is
represented only by the pole on sheet~II and a shifted pole on sheet~III.
The obtained pole clusters for the resonances are shown in Table~VII. 

\begin{widetext}
\begin{center}
\begin{table}[!htb]
\caption{The pole clusters for resonances in the $\sqrt{s}$ plane.
The poles corresponding to the $f_0^\prime(1500)$ on sheets III, V and VII are of
second order and that on sheet VI of  third order in our approximation.
~$\sqrt{s_r}\!=\!{\rm E}_r\!-\!i\Gamma_r/2$. \label{tab:clusters}}
\vspace*{0.1cm}
\def\arraystretch{1.3}
\begin{tabular}{cccccccc}
\hline
\hline ${\rm Sheet}$ & {} & $f_0(500)$ & $f_0(980)$ & $f_0(1370)$ & $f_0(1500)$ 
                          & $f_0^\prime(1500)$ & $f_0(1710)$ \\ \hline
II & {${\rm E}_r$} & $514.5\pm12.4$ & $1008.1\pm3.1$ & {} & {} & $1512.7\pm4.9$ & {} \\
{} & {$\Gamma_r/2$} & $465.6\pm5.9$ & $32.0\pm1.5$ & {} & {} & $285.8\pm12.9$ & {} \\
\hline 
III & {${\rm E}_r$} & $544.8\pm17.7$ & $976.2\pm5.8$ & $1387.6\pm24.4$ 
                    & {} & $1506.2\pm9.0$ & {} \\{}
& {$\Gamma_r/2$} & $465.6\pm5.9$ & $53.0\pm2.6$ & $166.9\pm41.8$ & {} & $127.9\pm10.6$ & {} \\
\hline 
IV & {${\rm E}_r$} & {} & {} & 1387.6$\pm$24.4 & {} & 1512.7$\pm$4.9 & {<} \\
{} & {$\Gamma_r/2$} & {} & {} & $178.5\pm37.2$ & {} & $216.0\pm17.6$ & {} \\
\hline 
V & {${\rm E}_r$} & {} & {} & $1387.6\pm24.4$ & $1493.9\pm3.1$ & $1498.9\pm7.2$ 
                            & $1732.8\pm43.2$ \\
{} & {$\Gamma_r/2$} & {} & {} & $260.9\pm73.7$ & $72.8\pm3.9$ 
                              & $142.2\pm6.0$ & $114.8\pm61.5$ \\
\hline VI & {${\rm E}_r$} & $566.5\pm29.1$ & {} & 1387.6$\pm$24.4 & $1493.9\pm5.6$
& $1511.4\pm4.3$ & 1732.8$\pm$43.2 \\
{} & {$\Gamma_r/2$} & $465.6\pm5.9$ & {} & $249.3\pm83.1$ & $58.4\pm2.8$ 
                    & $179.1\pm4.0$ & $111.2\pm8.8$ \\
\hline VII & {${\rm E}_r$} & $536.2\pm25.5$ & {} & {} & $1493.9\pm5.0$ 
                           & $1500.5\pm9.3$ & 1732.8$\pm$43.2 \\
{} & {$\Gamma_r/2$} & $465.6\pm5.9$ & {} & {} & $47.8\pm9.3$ & $99.7\pm18.0$ 
                    & $55.2\pm38.0$ \\
\hline 
VIII & {${\rm E}_r$} & {} & {} & {} & $1493.9\pm3.2$ & 1512.7$\pm$4.9 & 1732.8$\pm$43.2 \\
{} & {$\Gamma_r/2$} & {} & {} & {} & $62.2\pm9.2$ & $299.6\pm14.5$ & $58.8\pm16.4$ \\
\hline\hline
\end{tabular}
\end{table}
\end{center}
\end{widetext}

The obtained background parameters are shown in Table~\ref{tab:prefscenario}. 

\begin{table}
\begin{center}
\caption{Background parameters for the preferred scenario. \label{tab:prefscenario}}
\vspace*{0.1cm}
\def\arraystretch{1.5}
\begin{tabular}{ccc}
\hline\hline
$a_{11}$ & $a_{1\sigma}$ & $a_{1v}$ \\
0.0      & 0.0199        & 0.0    \\
\hline
$b_{11}$ & $b_{1\sigma}$ & $b_{1v}$ \\
0.0      & 0.0           & 0.0338   \\
\hline
$a_{21}$ & $a_{2\sigma}$ & $a_{2v}$ \\
$-$2.4649& $-$2.3222       & $-$6.611    \\
\hline
$b_{21}$ & $b_{2\sigma}$ & $b_{2v}$ \\
0.0    & 0.0             & 7.073     \\
\hline
$b_{31}$ & $b_{3\sigma}$ & $b_{3v}$ \\
0.6421   & 0.4851        & 0.0   \\
\hline
$s_{\sigma}$ & $s_{v}$ & \\ 
1.6338 GeV$^2$   & 
2.0857 GeV$^2$   & \\
\hline\hline
\end{tabular}
\end{center}
\end{table}

The small (zero for the elastic region) values of the $\pi\pi$-scattering background
parameters (obtained after allowing for the left-hand branch point at $s=0$) confirms
our assumption $S=S_B S_{res}$ and also that the representation of multichannel
resonances by the pole clusters on the uniformization plane is good and quite
sufficient.

It is important that we have practically obtained zero background for $\pi\pi$
scattering in the scalar-isoscalar channel because a reasonable
and simple description of the background should be a criterion for the correctness
of the approach. This shows that the consideration of the left-hand branch point
at $s=0$ in the uniformizing variable partly solves a problem of some approaches
(see, e.g., Ref.~\cite{Achasov-Shest}) where the wide-resonance parameters are
strongly controlled by the nonresonant background.

Another important conclusion in our approach is also related to a practically zero
background in $\pi\pi$-scattering: the contribution to the $\pi\pi$ scattering
amplitude from the crossed channels is given by allowing for the left-hand
branch point at $s=0$ in the uniformizing variable and the meson-exchange
contributions in the left-hand cuts. The zero background in the elastic-scattering
region is obtained only when taking into account the left-hand branch point in the
proper uniformizing variables both in the two-channel analysis of the processes
$\pi\pi\to\pi\pi,K\overline{K}$~\cite{SKN-epja02} and in the three-channel analysis of
the processes $\pi\pi\!\to\!\pi\pi,K\overline{K},\eta\eta$. This indicates that
the $\rho$- and $f_0(500)$-meson-exchange contributions in the left-hand cut
practically cancel each other.
One can show allowing for gauge invariance that the
vector- and scalar-meson exchanges contribute with opposite signs. Therefore, the
practically zero background in $\pi\pi$ scattering is an additional confirmation
that the $f_0(500)$ observed in the analysis as the pole cluster of type {\bf a} 
is indeed a particle (though very wide), not some dynamically formed resonance.
Therefore, one must consider at least in the background the coupled $\sigma\sigma$
channel which is not taken into account explicitly in the uniformizing variable
(\ref{w}). In this connection it is reasonable to interpret the effective
threshold at $s_\sigma=1.6338~{\rm GeV}^2$ in the background phase shift of the
$\pi\pi$-scattering amplitude as related to the $\sigma\sigma$ channel.
Only in this channel we have obtained a nonzero background phase shift in
$\pi\pi$ scattering ($a_{1\sigma}=0.0199$).

\end{document}